\documentclass[prd,amssymb,amsmath,amsfonts,superscriptaddress,nofootinbib,reprint,showpacs, longbibliography]{revtex4-1}

\usepackage{graphicx}
\usepackage{lmodern}
\usepackage{amsmath,amssymb}
\usepackage{mathrsfs}
\usepackage{amsfonts}
\usepackage[utf8]{inputenc}
\usepackage{url}
\usepackage[colorlinks]{hyperref}
\usepackage{xcolor}
\usepackage{multirow}
\usepackage[normalem]{ulem}
\usepackage{orcidlink}
\usepackage{placeins}
\usepackage{mathtools}
\usepackage[shortlabels]{enumitem}

\definecolor{dodgerblue}{HTML}{1E90FF}
\definecolor{viennared}{HTML}{DA0A14}
\definecolor{ctorange}{HTML}{FF6C0C}
\definecolor{wales}{HTML}{ff0038}
\definecolor{benettongreen}{HTML}{009421}
\definecolor{ferrarired}{HTML}{ff2800}
\definecolor{austriawienpurple}{HTML}{441678}

\DeclareFontFamily{OT1}{pzc}{}
\DeclareFontShape{OT1}{pzc}{m}{it}{<-> s * [1.10] pzcmi7t}{}
\DeclareMathAlphabet{\mathpzc}{OT1}{pzc}{m}{it}

\hypersetup{
     colorlinks=true,
     linkcolor=dodgerblue,
     filecolor=dodgerblue,
     citecolor = dodgerblue,      
     urlcolor=dodgerblue,
     }

\newcommand{\CIT}{TAPIR, California Institute of Technology, Pasadena, CA 91125, USA}
\newcommand{\LIGOLab}{LIGO Laboratory, California Institute of Technology, Pasadena, California 91125, USA}
\newcommand{\Amherst}{Department of Physics \& Astronomy, Amherst College, Amherst, Massachusetts 01002, USA}
\newcommand{\Cornell}{Cornell Center for Astrophysics and Planetary Science, Cornell University, Ithaca, New York 14853, USA}

\newcommand{\chip}{\chi_p}

\newcommand{\SEOB}{\texttt{SEOBNRv5EHM}{}}
\newcommand{\hlmNR}{$h^{\text{NR}}_{\ell m}$}
\newcommand{\hlmEOB}{$h^{\text{EOB}}_{\ell m}$}
\newcommand{\hNR}{$h^{\text{NR}}$}
\newcommand{\hEOB}{$h^{\text{EOB}}$}

\DeclarePairedDelimiterX{\norm}[1]{\lVert}{\rVert}{#1}

\begin{document}

\title{Revisiting the coprecessing frame in the presence of orbital eccentricity}

\author{Lucy M. Thomas \orcidlink{0000-0003-3271-6436}}
\email{lmthomas@caltech.edu}
\affiliation{\CIT}
\affiliation{\LIGOLab}

\author{Katerina Chatziioannou \orcidlink{0000-0002-5833-413X}}
\email{kchatziioannou@caltech.edu}
\affiliation{\CIT}
\affiliation{\LIGOLab}

\author{Sam Johar \orcidlink{0009-0003-5595-2228}}
\affiliation{\Amherst}
\affiliation{\LIGOLab}

\author{Taylor Knapp \orcidlink{0000-0001-8474-4143}}
\email{tknapp@caltech.edu}
\affiliation{\CIT}
\affiliation{\LIGOLab}

\author{Michael Boyle \orcidlink{0000-0002-5075-5116}}
\affiliation{\Cornell}

\date{\today}

\begin{abstract}

Accurate inclusion of both spin precession and orbital eccentricity effects in gravitational waveform models represents a key hurdle in our ability to fully characterize the properties of compact binaries. 
Virtually all efforts to model precession rely on a \emph{coprecessing} frame transformation, a time-dependent spatial rotation that tracks the dominant emission direction and simplifies the waveform morphology.
We assess the utility of the coprecessing frame transformation to separate out the effect of the precession of the orbital plane from the waveform in the presence of non-negligible orbital eccentricity.
We rely on $20$ numerical relativity simulations, which include the complete physical effects of spin precession and eccentricity in the strong-field, and compare waveforms in both the inertial and coprecessing frames.
Comparing against the eccentric, spin-aligned model \texttt{SEOBNRv5EHM}, we find that while the waveform \emph{mismatches} decrease in the coprecessing frame, they remain above the level required for accurate waveform modeling, $\sim 0.01$ or higher for large inclinations.
Further improvements, e.g., modeling mode asymmetries as already pursued for quasicircular binaries, will likely prove essential. 
We also find that by removing the dominant amplitude and phase modulations from the waveform, the coprecessing frame facilitates surrogate modeling, achieving lower errors at a fixed number of basis elements compared to the inertial frame. 
Our results demonstrate both the utility and the limitations of the coprecessing frame as a cornerstone in waveform modeling for eccentric and precessing binaries.
\end{abstract}

\maketitle 
\section{Introduction}
\label{sec:Introduction}

Detection of gravitational waves (GWs) from the coalescence of binary black holes (BBHs) has now become routine, with over 200 such events observed to date \cite{LIGOScientific:2025slb, Wadekar:2023gea,Nitz:2021zwj}.
While the orbits of these binaries are expected to efficiently circularize through the emission of GWs as they inspiral towards the merger~\cite{Peters:1963ux,Peters:1964zz}, the detection of a binary with non-zero eccentricity would be a key tracer of formation in a dynamical environment \cite{Lower:2018seu,Zevin:2021rtf}. 
Hints of non-zero eccentricity in BBH signals \cite{Gayathri:2020coq,Gamba:2021gap,Iglesias:2022xfc,Gupte:2024jfe,Planas:2025jny,Romero-Shaw:2025vbc} and a neutron star-BH system \cite{Morras:2025xfu,Jan:2025fps} have been recovered from the data.
However, modeling of the BBH signals is incomplete as it neglects the effect of spin precession, arising from interactions between the orbital and spin angular momenta~\cite{Apostolatos:1994mx}.
Indeed, binaries which form in a dynamical environment have random spin directions and thus spin precession is expected together with any remaining orbital eccentricity.
Moreover, eccentricity and spin precession effects can appear degenerate in GW data \cite{Ramos-Buades:2019uvh,CalderonBustillo:2020xms,Tibrewal:2026jci,RoyChowdhury:2026xgb}, especially close to merger and for binaries observed face-on \cite{Romero-Shaw:2020thy, Gayathri:2020coq, Romero-Shaw:2022fbf}, though this degeneracy is dependent on many factors such as the separation between orbital and precession timescales \cite{Morras:2025nlp}. 
Additionally, eccentricity is coupled to the spin evolution \cite{Gergely:1998sr,Memmesheimer:2004cv,Konigsdorffer:2005sc,Klein:2010ti,Cornish:2010cd,Levin:2010dt,Gopakumar:2011zz,Tessmer:2014vha,Fumagalli:2023hde, Henry:2023tka,Phukon:2025yva}, further necessitating accurate waveform models with both effects.

The above considerations motivate development of eccentric waveform models \cite{Gamba:2024cvy,Gamboa:2024hli, Nagar:2024oyk, Paul:2024ujx, Ramos-Buades:2026kbq}, with extensions from aligned to precessing spins forming the current frontier of modeling efforts. 
An added complication of eccentric modeling is that there is no gauge-independent definition of eccentricity in the strong-field regime, so different definitions are used internally between models, obfuscating comparison between models and with numerical relativity (NR) simulations~\cite{Scheel:2025jct,Ferguson:2023vta,Healy:2020vre} or astrophysical population models \cite{Vijaykumar:2024piy}. 
Current models that include both eccentricity and precession are either inspiral-only \cite{Morras:2025nlp}, or impose circularization before merger \cite{Liu:2023ldr,Albanesi:2025txj}.

Eccentric and precessing waveform models will likely rely on a key element currently used by all precessing models: the coprecessing frame approximation \cite{Schmidt:2012rh,OShaughnessy:2012lay}. 
The coprecessing frame is a time-dependent coordinate transformation which effectively tracks the precession of the orbital plane and keeps the $z$-axis aligned with the instantaneous direction of maximum emission. 
This method breaks the complex and feature-rich behaviour of a precessing waveform into two simpler pieces: an approximately non-precessing waveform, and a time-dependent rotation which represents the transformation to the coprecessing frame. 
The two pieces are then modelled separately, and combined to produce the full precessing waveform as observed by a detector in an inertial frame. 

While the coprecessing frame plays a role in all current models, that role is different between semianalytic and surrogate models.
Precessing semianalytic waveform models such as Phenomenological (Phenom) \cite{Schmidt:2012rh, Hamilton:2025xru} and Effective-One-Body (EOB) \cite{Pan:2013rra, Gamba:2021ydi, Ramos-Buades:2023ehm} families begin from the assumption that the waveform in the coprecessing frame can be physically identified with an approximately non-precessing one. 
More recently, additional effects to account for the differences between spin-aligned and precessing waveforms in the coprecessing frame have been included, such as asymmetries between positive and negative $m$-modes \cite{Boyle:2014ioa,Thompson:2023ase, Estelles:2025zah}, and changes to the ringdown frequencies and final spin of the remnant \cite{Hamilton:2023znn}. 
This is in contrast to NR surrogate models, which use the coprecessing frame as a decomposition tool to simplify the waveform components to be modelled.
Each component is modelled in a data-driven way with no assumption about their morphology, or that coprecessing frame waveforms resemble spin-aligned ones.

The continued usefulness of the coprecessing frame in the presence of orbital eccentricity is an open question.
\citet{Gamba:2024cvy} demonstrated that the presence of eccentricity does not preclude the definition of the coprecessing frame transformation.
Furthermore, they inspected the modes of 1 eccentric and precessing simulation and showed that they resemble those of the equivalent eccentric and spin-aligned system after being transformed to the coprecessing frame.
\citet{Shaikh:2025tae} further showed that the eccentricity defined in the coprecessing frame is similar to that of the equivalent spin-aligned system.

With the modeling applications in mind, in this work we revisit the coprecessing frame and study it in detail in the presence of orbital eccentricity. 
First, we consider how faithfully eccentric, precessing waveforms resemble eccentric, spin-aligned ones when transformed into the coprecessing frame. 
We compare NR waveforms from the SXS catalog~\cite{Scheel:2025jct} that contain eccentricity and precession to \SEOB~\cite{Gamboa:2024hli} that is restricted to aligned spins in Sec.~\ref{sec:ComparisonsintheCoprecessingFrame}. 
By comparing the NR waveforms in both the inertial and coprecessing frame to \SEOB, we quantify the gain from transforming to the coprecessing frame, and whether more physics is needed to fully account for spin precession also in the coprecessing frame.
Second, for surrogate models, which will inherit the complete strong-field spin precession and eccentricity effects from NR, the more pertinent question is not one of missing physics, but whether the coprecessing frame facilitates modeling. 
In Sec.~\ref{sec:BuildingReducedBases} we investigate whether using the coprecessing frame for eccentric and precessing simulations leads to smoother and more slowly-varying waveform components (amplitudes and phases), and therefore more accurate representations of the parameter space with smaller basis sizes. 
Overall, we find that the coprecessing frame, despite some limitations, will remain essential for precession modeling even in the presence of eccentricity. 
We present our conclusions in Sec.~\ref{sec:conclusions}.

\section{Methodology}
\label{sec:Methods}

In this section, we introduce the methodology used in this work, including the coprecessing frame in Sec.~\ref{subsec:CoprecFrame}, the NR simulations considered in Sec.~\ref{subsec:NRSims}, and the procedure for computing mismatches between the NR simulations and \SEOB~in Sec.~\ref{subsec:MismatchProcedure}.

\subsection{Coprecessing Frame}
\label{subsec:CoprecFrame}

\begin{figure*}[]
    \centering
    \includegraphics[width=1.8\columnwidth]{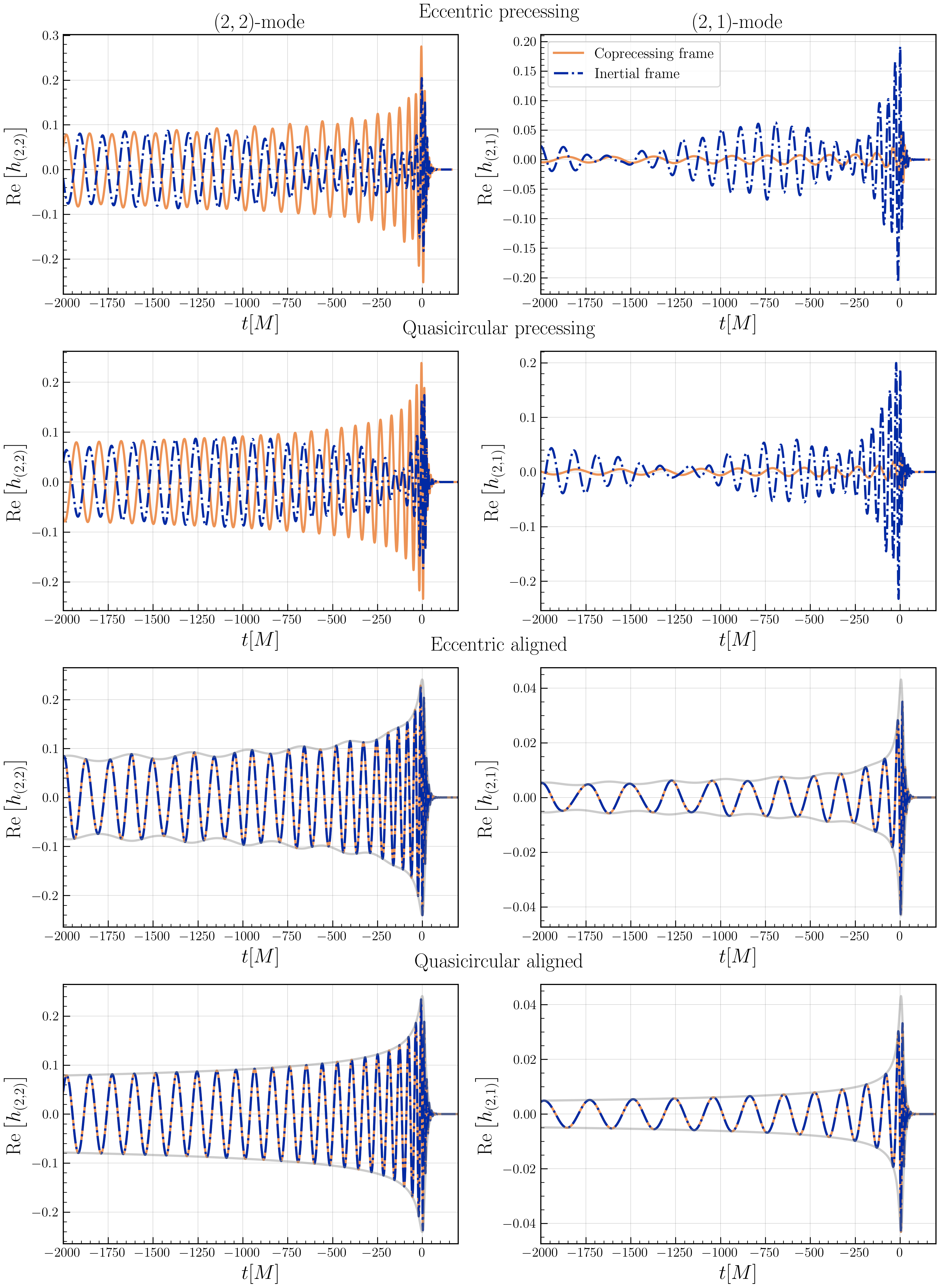}
    \caption{Effect of the coprecessing frame on waveform modes, the $(2,2)$ mode (left column) and $(2,1)$ mode (right column), for four different binary configurations: eccentric and precessing (top row), precessing and quasicircular (second row), non-precessing and eccentric (third row), and quasicircular and non-precessing (bottom row). In the non-precessing cases, the coprecessing frame leaves the waveform modes identical, but for the top two panels it simplifies the precession modulations and restores the mode hierarchy. The binary parameters are that of the SXS waveform \texttt{SXS:BBH:4286}, with the full NR waveform shown in the top row, while the other three rows use waveforms with the same parameters but removing the in-plane spins or eccentricity as appropriate, generated using either \texttt{SEOBNRv5PHM} \cite{Ramos-Buades:2023ehm} or \texttt{SEOBNRv5EHM} \cite{Gamboa:2024hli}. In the aligned spin cases we also plot the waveform mode envelopes, to show the effect of the eccentricity on the waveform. Waveforms are plotted as a function of time in units of the total mass $M$.}
    \label{fig:coprec_frame}
\end{figure*}

In this subsection, we describe the \emph{coprecessing frame} and examine its effect on waveforms with both eccentricity and precession in varying combinations, to serve as intuition for the subsequent analysis.
In Fig.~\ref{fig:coprec_frame} and throughout this work, we use the \texttt{scri} package \cite{Boylescri,Boyle:2015nqa} to perform rotations into the coprecessing frame. 
As described below, we use the waveform modes to identify the time-dependent direction of principal GW emission, and then rotate the waveform to this frame. 
Further details and the exact mathematical definition can be found in Ref.~\cite{Boyle:2014ioa}.

In a binary where the spins are aligned with the orbital angular momentum and there is no spin precession, the amplitude and frequency of the signal increases monotonically throughout the inspiral towards merger. 
The strain can be decomposed into a set of spin-2 weighted spherical harmonics,
\begin{equation}
    h(t, \iota, \phi; \vec{\Theta}) = \sum\limits_{\ell, m} h_{\ell m}(t;\vec{\Theta}) Y^{-2}_{\ell m}(\iota,\phi)\,,
    \label{eq:swsh}
\end{equation}
where $\vec{\Theta}$ are the intrinsic parameters of the binary (the masses, $m_1$ and $m_2$, and spins, $\vec{\chi}_1$ and $\vec{\chi}_2$), and $\iota,\phi$ are the polar and azimuthal angles on the unit sphere respectively. 
Then, the $(\ell=2,m=2)$ mode remains dominant throughout the inspiral, and there is an exact symmetry between the positive and negative-$m$ modes,
\begin{equation}
    h_{\ell m} = (-1)^{\ell}h^{*}_{\ell -m}\,,
\label{eq:symmetry}
\end{equation}
where $*$ represents the complex conjugate.

In a spin precessing system, couplings between the orbital angular momentum and the spin angular momenta cause the orbital plane and spin vectors to precess and affect the binary dynamics \cite{Apostolatos:1994mx,Kidder:1995zr}.
Precession complicates the signal by introducing time-dependent amplitude and phase modulations, transferring power into higher-order multipoles through mode-mixing and altering their frequency, introducing a time-dependence to binary inclination, and breaking the symmetry in Eq.~\eqref{eq:symmetry}. 
After the merger, it also modifies the ringdown of a remnant black hole \cite{Finch:2021iip, Zhu:2023fnf}, as well as the remnant spin and recoil velocity \cite{Buonanno:2007sv,Boyle:2007ru,Rezzolla:2007rz,Campanelli:2007ew,Lousto:2012gt, Zlochower:2015wga,Hofmann:2016yih,Ramos-Buades:2020noq}.
Since the root cause of most of these effects is the precessional binary motion, the coprecessing frame is effectively designed to track the motion of the orbital plane by aligning its $z$-axis with the principal direction of emission \cite{OShaughnessy:2011pmr,Schmidt:2012rh}. 
The effects of precession on the observed waveform are reduced in this time-dependent frame, as can be seen in Fig.~\ref{fig:coprec_frame} in the second-from-top panel. 

In Fig.~\ref{fig:coprec_frame}, we plot the $(2,2)$-mode (left) and  $(2,1)$-mode (right) (the second-from-top panel is generated using the \texttt{SEOBNRv5PHM} model \cite{Ramos-Buades:2023ehm}) for a spin-precessing binary with $q\equiv m_1/m_2=4$, $\vec{\chi}_1=\left[0.72, 0.61, 0.07\right]$, $\vec{\chi}_2=\left[0.17,0.46, 0.1\right]$ in both the inertial frame (blue), and after being transformed to the coprecessing frame (orange). 
The time-dependent amplitude modulations, which are clearly visible in the inertial frame, are all but gone in the coprecessing frame, and the non-precessing mode hierarchy of the $(2,2)$ mode being dominant is restored. 
Mode-mixing is more evident in the $(2,1)$ mode frequency which is expected to be proportional to $m$~\cite{Blanchet:2002av}.
This is true in the coprecessing frame, but not the inertial frame where the $(2,2)$ and $(2,1)$ modes have the same frequency.
Because mode-mixing between the $(2,2)$ and $(2,1)$-modes is effectively removed, the frequency of the $(2,1)$-mode oscillations is reduced by a factor of $2$ in the coprecessing frame as it no longer contains contributions from the inertial frame $(2,2)$-mode.
Comparing this to the bottom panel, which shows the equivalent non-precessing modes with the same binary parameters but with the in-plane spin components set to zero, the coprecessing frame transformation now has no effect on the waveform.
Since there is no spin precession, the waveform modes are almost identical to the coprecessing frame ones in the second-from-top panel. 
While the coprecessing frame removes the dominant effect in the waveform of the precession of the orbital plane of the binary, there is however no frame rotation (time-dependent or otherwise) which can restore the symmetry of Eq.~\eqref{eq:symmetry} \cite{Boyle:2014ioa}, or capture higher-order spin couplings.

Formally, the coprecessing frame is defined by a time-dependent rotation $R(t)\in \rm{SO}(3)$. 
This rotation may be parameterized by Euler angles $\lbrace \alpha(t), \beta(t), \gamma(t) \rbrace$ \cite{Boyle:2011gg,Schmidt:2012rh,OShaughnessy:2012lay},  such that the 
$z$-axis approximately tracks a preferred radiation axis, commonly identified as the dominant quadrupolar emission \cite{Schmidt:2010it,OShaughnessy:2011pmr}. 
The spherical harmonic modes as defined in Eq.~\eqref{eq:swsh} transform between the inertial and coprecessing frames according to
\begin{equation}
    h_{\ell m} (t) = \sum^{\ell}_{m' = -\ell} D^{\ell}_{m m'}\left(\alpha(t),\beta(t),\gamma(t)\right) \tilde{h}_{\ell m'}(t)\,,
    \label{eq:wignerd}
\end{equation}
where $D^{\ell}_{m m'}$ are Wigner $D$-matrices, $h_{\ell m}(t)$ are the waveform modes in the inertial frame, and $\tilde{h}_{\ell m}(t)$ in the coprecessing frame. 
The remaining freedom in 
$\gamma(t)$ is fixed by imposing the minimal-rotation condition,
\begin{equation}
\dot{\gamma}(t) = -\dot{\alpha}(t) \cos\left(\beta(t)\right)\,,
\label{eq:minimalrotation}
\end{equation}
which ensures that the coprecessing frame undergoes no instantaneous rotation about its own $z$-axis beyond that required to track the radiation axis \cite{Buonanno:2002fy,Boyle:2011gg}. 
In this minimally rotating coprecessing frame, the waveform modes 
$\tilde{h}_{\ell m}(t)$ exhibit suppressed precession-induced amplitude and phase modulations and evolve on the slower radiation-reaction timescale. 
This construction yields a unique and smooth separation between intrinsic binary dynamics and purely kinematic precessional effects, providing a mathematically well-defined foundation for modeling and interpolating generic precessing signals.

We now additionally consider the effect of eccentricity in the waveform, and plot in the top panel of Fig.~\ref{fig:coprec_frame} the equivalent waveform with additionally an eccentricity of $e=0.1$ (this corresponds to the SXS simulation \texttt{SXS:BBH:4286}~\cite{Scheel:2025jct}).\footnote{Here we quote the eccentricity provided in the simulation metadata. We return to the issue of eccentricity definition in the next subsection.} 
Similarly to the quasicircular precessing case, the amplitude modulations in the $(2,2)$- and $(2,1)$-modes are removed in the coprecessing frame.
The resulting modes closely resemble those of the equivalent eccentric but non-precessing waveform, shown in the third panel. 
In this non-precessing eccentric case (waveforms were produced with the \texttt{SEOBNRv5EHM} model \cite{Gamboa:2024hli}), the transformation to the coprecessing frame again leaves the waveform identical.
Comparing the bottom two rows, the effect of the orbital eccentricity in either frame is encoded in the difference between peak amplitude between apastron and periastron passages, as highlighted by the amplitude envelopes plotted in grey.

\subsection{Numerical Relativity Simulations}
\label{subsec:NRSims}

\begin{table*}[]
    \centering
    \begin{tabular}{c|ccccccc}
    \hline
    \hline
    Sim ID & Mass ratio & Primary spin & Secondary spin & \multirow{2}{*}{$\chi_{\text{eff}}$} & \multirow{2}{*}{$\chi_{\text{p}}$} & Eccentricity & Number of \\
     & $q$ & $\vec{\chi}_{1}$ &  $\vec{\chi}_{2}$ &  &  & $e^{\text{NR}}_{\text{gw}}$ & orbits \\
    \hline
    \texttt{SXS:BBH:0078} & $1.50$ & $\left[0.38, 0.32, 0.02\right]$ &  $\left[0.00, 0.00, 0.00\right]$ & $0.00$ & $0.49$ & $0.020$ & $20.7$ \\
    \texttt{SXS:BBH:0082} & $1.50$ & $\left[0.35, 0.35, 0.03\right]$ &  $\left[0.00, 0.00, 0.00\right]$ & $0.00$ & $0.50$ & $0.020$ & $17.2$ \\
    \texttt{SXS:BBH:0088} & $1.00$ & $\left[0.49, 0.09, 0.00\right]$ &  $\left[0.00, 0.00, 0.00\right]$ & $0.00$ & $0.50$ & $0.058$ & $31.3$ \\
    \texttt{SXS:BBH:0094} & $1.50$ & $\left[0.37, 0.34, 0.02\right]$ &  $\left[0.00, 0.00, 0.00\right]$ & $0.00$ & $0.50$ & $0.012$ & $18.9$ \\
    \texttt{SXS:BBH:0097} & $1.50$ & $\left[0.34, 0.36, 0.03\right]$ &  $\left[0.00, 0.00, 0.00\right]$ & $0.00$ & $0.50$ & $0.035$ & $16.6$ \\
    \texttt{SXS:BBH:0098} & $1.50$ & $\left[0.39, 0.31, 0.02\right]$ &  $\left[0.00, 0.00, 0.00\right]$ & $0.00$ & $0.50$ & $0.019$ & $25.0$ \\
    \texttt{SXS:BBH:0179} & $1.50$ & $\left[0.00, 0.00, 0.99\right]$ &  $\left[0.07, 0.12, 0.14\right]$ & $0.65$ & $0.09$ & $0.028$ & $21.2$ \\
    \texttt{SXS:BBH:1160} & $3.00$ & $\left[0.25, 0.51, 0.41\right]$ &  $\left[0.39, -0.35, 0.60\right]$ & $0.22$ & $0.60$ & $0.025$ & $19.4$ \\
    \texttt{SXS:BBH:1388} & $3.00$ & $\left[0.25, 0.50, 0.42\right]$ &  $\left[-0.21, 0.08, 0.56\right]$ & $0.42$ & $0.57$ & $0.045$ & $20.6$ \\
    \texttt{SXS:BBH:3710} & $1.05$ & $\left[0.27, 0.57, 0.18\right]$ &  $\left[-0.16, -0.88, 0.38\right]$ & $0.27$ & $0.84$ & $0.097$ & $14.2$ \\
    \texttt{SXS:BBH:3711} & $1.37$ & $\left[0.11, -0.58, 0.74\right]$ &  $\left[-0.05, -0.05, -0.37\right]$ & $0.21$ & $0.72$ & $0.078$ & $11.9$ \\
    \texttt{SXS:BBH:3712} & $1.25$ & $\left[0.16, -0.29, -0.04\right]$ &  $\left[-0.05, 0.34, 0.73\right]$ & $0.30$ & $0.33$ & $0.047$ & $13.0$ \\
    \texttt{SXS:BBH:3713} & $2.00$ & $\left[0.60, -0.35, 0.59\right]$ &  $\left[-0.54, -0.12, 0.50\right]$ & $0.47$ & $0.83$ & $0.049$ & $12.0$ \\
    \texttt{SXS:BBH:3714} & $5.32$ & $\left[0.50, 0.38, -0.02\right]$ &  $\left[0.13, -0.26, 0.41\right]$ & $0.00$ & $0.62$ & $0.04$ & $16.6$ \\
    \texttt{SXS:BBH:3715} & $2.29$ & $\left[0.64, -0.56, 0.20\right]$ &  $\left[-0.25, -0.02, -0.48\right]$ & $-0.13$ & $0.87$ & $0.019$ & $8.6$ \\
    \texttt{SXS:BBH:3716} & $4.85$ & $\left[0.60, 0.22, 0.37\right]$ &  $\left[-0.52, -0.08, 0.07\right]$ & $0.27$ & $0.67$ & $0.029$ & $18.3$ \\
    \texttt{SXS:BBH:3717} & $2.40$ & $\left[-0.60, 0.09, 0.65\right]$ &  $\left[-0.14, 0.71, 0.25\right]$ & $0.46$ & $0.74$ & $0.071$ & $15.8$ \\
    \texttt{SXS:BBH:3718} & $1.19$ & $\left[-0.24, -0.06, 0.36\right]$ &  $\left[0.63, 0.62, 0.13\right]$ & $0.25$ & $0.73$ & $0.087$ & $14.5$ \\
    \texttt{SXS:BBH:3722} & $3.82$ & $\left[0.85, 0.03, 0.42\right]$ &  $\left[0.52, -0.59, 0.22\right]$ & $0.33$ & $0.90$ & $0.051$ & $8.4$ \\
    \texttt{SXS:BBH:3974} & $1.00$ & $\left[0.30, -0.74, 0.01\right]$ &  $\left[0.00, 0.00, 0.00\right]$ & $0.00$ & $0.80$ & $0.050$ & $16.7$ \\
    \texttt{SXS:BBH:4286} & $4.00$ & $\left[0.72, 0.61, 0.07\right]$ &  $\left[0.17, 0.46, 0.10\right]$ & $0.00$ & $0.95$ & $0.002$ & $32.8$ \\
    \texttt{SXS:BBH:4290} & $12.00$ & $\left[-0.67, 0.44, 0.04\right]$ &  $\left[0.00, 0.00, 0.00\right]$ & $0.03$ & $0.80$ & $0.002$ & $42.6$ \\
    \hline
    \hline
    \end{tabular}
    \caption{Parameters of the 22 NR simulations selected for this study. The mass ratio $q$, primary and secondary spins $\vec{\chi}_{1}$, $\vec{\chi}_{2}$, and effective spins $\chi_{\text{eff}}$, $\chi_{p}$ are taken from the simulation metadata and quoted at the reference time of the simulation after junk radiation. Spin vectors are defined in a frame where the $z$-axis is aligned with the orbital angular momentum, and the more massive BH is on the positive $x$-axis.
    The number of orbits is from the reference time until the formation of a common horizon, and is also taken from the simulation metadata. The eccentricity $e^{\text{NR}}_{\text{gw}}$ is measured from the waveform itself using the \texttt{gw\_eccentricity} package, with the procedure described in Secs.~\ref{subsec:NRSims} and~\ref{subsec:MismatchProcedure}. The last two simulations are excluded from the mismatch calculations of Sec.~\ref{sec:ComparisonsintheCoprecessingFrame} because their $e^{\text{NR}}_{\text{gw}}$ fails our eccentricity threshold of $0.01$. }
    \label{tab:NRSimsParams}
\end{table*}

We assess the coprecessing frame with NR simulations taken from the most recent publicly-available SXS catalogue \cite{Scheel:2025jct}. 
We choose eccentric and precessing simulations based on the values of parameters at the reference time of the simulation, as provided in the simulation metadata. 
As described in Sec. 3.5 of Ref.~\cite{Scheel:2025jct}, the reference time is set to the relaxation time of the simulation, which is the time at which junk radiation has decayed as measured by high frequency oscillations in the waveform. 
For the eccentricity, we place a threshold of $e>0.01$, and for the precessing spins, we place a minimum threshold that at least one of the BHs has an in-plane spin value $\chi_{i\perp}>0.01$. 
Additionally, we require simulations to have at least $8$ inspiral orbits.
This results in 22 eccentric and precessing simulations, the parameters of which are displayed in Table~\ref{tab:NRSimsParams}. 
Eccentricities are measured using one of two different methods as described in Sec.~3.4 of Ref.~\cite{Scheel:2025jct} which use an inversion of the Kepler equation at 1PN (either approximately analytically or numerically) to solve for the eccentric anomaly as a function of time. 
Both methods are initially used for each simulation with eccentricity $e>0.01$, and then the result with the smallest value of the $L2$ norm of the fit residual divided by the number of fit parameters is selected.
These eccentricity values are only used for choosing the simulations.
Subsequent studies rely on an eccentricity redefinition using the procedure described below. 
However, the different definitions will agree in the limit of $e\rightarrow0$  \cite{Shaikh:2023ypz}.

In order to prepare the NR simulations for comparison to EOB waveforms, first we remove the junk radiation by cutting off at the reference time, as specified by the simulation metadata. 
We then ascertain a merger time as in Ref.~\cite{Gamboa:2024hli}, by calculating the frame-invariant amplitude $\left|h\right|$ at each time $t$,
\begin{equation}
        \left|h\right|(t) = \sqrt{\sum_{\ell, |m|\leq\ell} \left|h_{\ell m}\right|^2(t)}\,,
        \label{eq:hamp}
\end{equation}
and define the merger time $t_{c}$ to be the time at which $\left|h\right|$ is maximized.\footnote{In Ref.~\cite{Gamboa:2024hli} a slightly different definition of $t_c$ was used to account for binaries with large eccentricities possibly having large amplitude peaks in the late inspiral at close periastron passages. We do not adopt this modification as the eccentricities considered in this work are moderate.}
We then time-shift the waveform such that zero corresponds with $t_{c}$. 
As we are mainly interested in the impact of the coprecessing frame on the inspiral, we also cut the post-merger portion of the waveform, after $t_c$.

We redefine the eccentricity of a simulation from the waveform itself using the \texttt{gw\_eccentricity} package \cite{Shaikh:2025tae}. 
Specifically, we calculate the GW frequency $\omega_{\text{gw}}$ defined from the $(2,2)$- and $(2,-2)$-mode frequencies in the coprecessing frame (Eq.~(8) of Ref.~\cite{Shaikh:2025tae}):
\begin{equation}
    \omega_{\text{gw}} = \frac{d\phi_{\text{gw}}}{dt} = \frac{1}{2}\left(\omega^{\text{copr}}_{2,2}-\omega^{\text{copr}}_{2,-2}\right)\,.
\end{equation}
We then calculate the eccentricity and mean anomaly as:
\begin{equation}
    e_{{\text{gw}}}(t) = \frac{\sqrt{\omega^{p}_{\text{gw}}(t)} - \sqrt{\omega^{a}_{\text{gw}}(t)}}{\sqrt{\omega^{p}_{\text{gw}}(t)} + \sqrt{\omega^{a}_{\text{gw}}(t)}}\,,
    \label{eq:e_gw}
\end{equation}
\begin{equation}
    l_{\text{gw}}(t) = 2\pi \frac{t - t^{p}_{i}}{t^{p}_{i+1} - t^{p}_{i}}\,.
    \label{eq:l_gw}
\end{equation}
Here, the $p$ and $a$ superscripts refer to periastron and apastron respectively, and the $i$ subscripts refer to successive periastron passages. The functions of time $\omega^{a}_{\text{gw}}(t)$ and $\omega^{p}_{\text{gw}}(t)$ represent smooth interpolants through the values of the GW frequency through the apastrons and periastrons respectively.
Before identifying successive periastron and apastron passages, we also remove the secular trend in the waveform amplitude using the in-built \texttt{AmplitudeFits} method which fits a power-law model inspired by PN expressions in the quasicircular limit~\cite{Shaikh:2023ypz}. 

The eccentricity and GW frequency constructed here are in a sense orbit-averaged approximations, as the interpolants are smoothly varying between the periastron and apastron in Eq.~\eqref{eq:e_gw}.
They thus might not represent the full dynamic variation on timescales shorter than the orbital timescale. 

\subsection{Mismatch Calculation}
\label{subsec:MismatchProcedure}

In order to find the \SEOB~waveform which best matches a given NR simulation, we compute mismatches between the NR waveform strain, denoted by \hNR, and EOB strains, denoted by \hEOB. 
We compute white-noise mismatches throughout and so do not set a total mass or luminosity distance scale. 
Instead, we use the entire length of each NR inspiral waveform, from the start time at which we define the initial eccentricity (described below) up to $t_c$.
Our mismatch calculation proceeds as follows, and is similar to the procedure described in Ref.~\cite{Gamboa:2024hli}.

First, we preprocess the NR simulations as described in Sec.~\ref{subsec:NRSims}. 
We remove the junk radiation, shift the times such that $t_c=0$, remove the postmerger, and extract the eccentricity $e_{\text{gw}}(t)$, mean anomaly $l_{\text{gw}}(t)$, and orbit-averaged GW frequency $\omega_{\text{gw}}(t)$ from the waveform itself as a function of time. 
As our initial time for the lower integration limit of the mismatch, we choose the first apastron after the beginning of the interpolants for $\omega^{\text{p}}_{\text{gw}}(t)$ and $\omega^{\text{a}}_{\text{gw}}(t)$. 
In practice, this cuts at most $2$ orbits from the waveform in addition to the junk radiation. 
The first portion cut (maximum $1.5$ orbits) is prior to the beginning of the interpolants, because Eqs.~\eqref{eq:e_gw} and~\eqref{eq:l_gw} require at least a full orbit (one apastron and one periastron) to identify minima and maxima of $\omega_{\text{gw}}$. The next half orbit ensures that we specify the initial parameters at apastron, though it may not be needed if the binary passes through a periastron first, then an apastron.
We start at apastron to avoid generation of \SEOB~waveforms with initial conditions close to periastron, which (for high eccentricities) may be well into the strong field regime and prove difficult for the initial conditions solver. 
 We denote the values of the eccentricity, orbit-averaged GW frequency, and mean anomaly at this start time by explicitly dropping the dependence on time, i.e., $e^{\text{NR}}_{\text{gw}}$, $\omega^{\text{NR}}_{\text{gw}}$, and $l^{\text{NR}}_{\text{gw}}=\pi$, where the superscript denotes that these are measured from the NR waveform.\footnote{Two NR simulations that passed the thresholds of Sec.~\ref{subsec:NRSims} of eccentricity $e>0.01$ nonetheless have $e^{\text{NR}}_{\text{gw}}<0.01$. We exclude them from subsequent calculations.}

When computing mismatches between the NR and EOB waveforms, we use the same values for mass ratio $q$, and aligned spin values $\chi_{i\parallel}$ for both waveforms, and we also assume that the mean anomaly is the same, $l^{\text{EOB}}_{\text{gw}} = l^{\text{NR}}_{\text{gw}}=\pi$.
In this final assumption, we utilize the fact that despite the differing definitions of anomaly parameter between the NR simulations and \SEOB~(the latter uses a relativistic anomaly input), these two definitions align at apastron or periastron.
We optimize mismatches over the eccentric parameters of the \SEOB~waveform, $e^{\text{EOB}}_{\text{gw}}$ and $\omega^{\text{EOB}}_{\text{gw}}$, as well time $t_0$ and phase $\phi_0$ shifts. 
These parameters over which we optimize: $\lbrace e^{\text{EOB}}_{\text{gw}},\omega^{\text{EOB}}_{\text{gw}},t_0,\phi_0\rbrace$, in addition to those we hold fixed: $\lbrace q, \chi_{i\parallel},l_{\text{gw}}=\pi, \iota\in\left[0, \pi/4, \pi/2\right]\rbrace$, as well as those which are scaled out or not utilized for white noise dimensionless strain: $\lbrace M, d_L, \psi, \alpha, \delta \rbrace$\footnote{Here $M$ is the total mass of the binary (in units of solar masses), $d_L$ is the luminosity distance, $\psi$ is the polarization angle, and the sky location is parameterized by right ascension $\alpha$ (not to be confused with the precession Euler angle $\alpha(t)$) and declination $\delta$.}, make up the $14$ parameters for an eccentric spin-aligned BBH waveform starting from an initial frequency.

The mismatch optimization is summarized in the following equation:
\begin{equation}
\mathcal{MM}=1-\max\limits_{e^{\text{EOB}}_{\text{gw}},\omega^{\text{EOB}}_{\text{gw}},t_0,\phi_0} \left[\langle \hat{h}^{\text{EOB}}, \hat{h}^{\text{NR}} \rangle|_{q,\chi_{i\parallel},l_{\text{gw}}=\pi}\right]\,,
\end{equation}
where
\begin{equation}
\langle h_1, h_2 \rangle = 4\mathcal{R}\int^{f_{\text{max}}}_{f_{\text{min}}} \frac{\tilde{h}_1(f) \tilde{h}^*_2(f)}{S_n(f)}\, df\,,
\label{eq:mismatch}
\end{equation}
is the usual inner product where we set $S_n(f)\sim1$, and 
\begin{equation}
    \hat{h} = \frac{h}{\sqrt{\langle h, h \rangle}}\,.
\end{equation}
In practice, we define a grid of $101$ possible eccentricities $\mathcal{I}_{e}=\left[e^{\text{NR}}_{\text{gw}}+\delta e, e^{\text{NR}}_{\text{gw}} - \delta e\right]$ and take $\delta e=e^{\text{NR}}_{\text{gw}}/5$. 
For each $e\in\mathcal{I}_{e}$, we then find the value of $\omega^{\text{EOB}}_{\text{gw}}$ that minimizes the difference in time to merger between \hlmNR and \hlmEOB, using repeated \SEOB~evaluations with a \texttt{SciPy} optimizer.
Then for each value of $e\in\mathcal{I}_{e}$, we compute the mismatch between the corresponding \SEOB~waveform, optimizing numerically over time shifts, taking the maximum time shift $t_0$ to be $10\%$ of the length of the waveform, and analytically optimizing over phase shift $\phi_0$ \cite{Capano:2013raa}. 
We choose to search over grids for optimal values of $e^{\text{EOB}}_{\text{gw}}$ and $\omega^{\text{EOB}}_{\text{gw}}$ rather than using, for example a peak finder, for simplicity and easy parallelization of computation.
We validated our implementation of the above calculation by reproducing the distribution of mismatches between \SEOB\, and eccentric, spin-aligned NR simulations reported in Ref.~\cite{Gamboa:2024hli}.

We compute mismatches between \SEOB~and the NR simulations, both with and without the latter being transformed into the coprecessing frame. 
We do this for the full waveform strain with inclination angles $\iota\in\left[0,\pi/4, \pi/2\right]$. 

\section{Physical Meaningfulness of Coprecessing Frame Waveform}
\label{sec:ComparisonsintheCoprecessingFrame}

\begin{figure*}[]
    \centering
    \includegraphics[width=\textwidth]{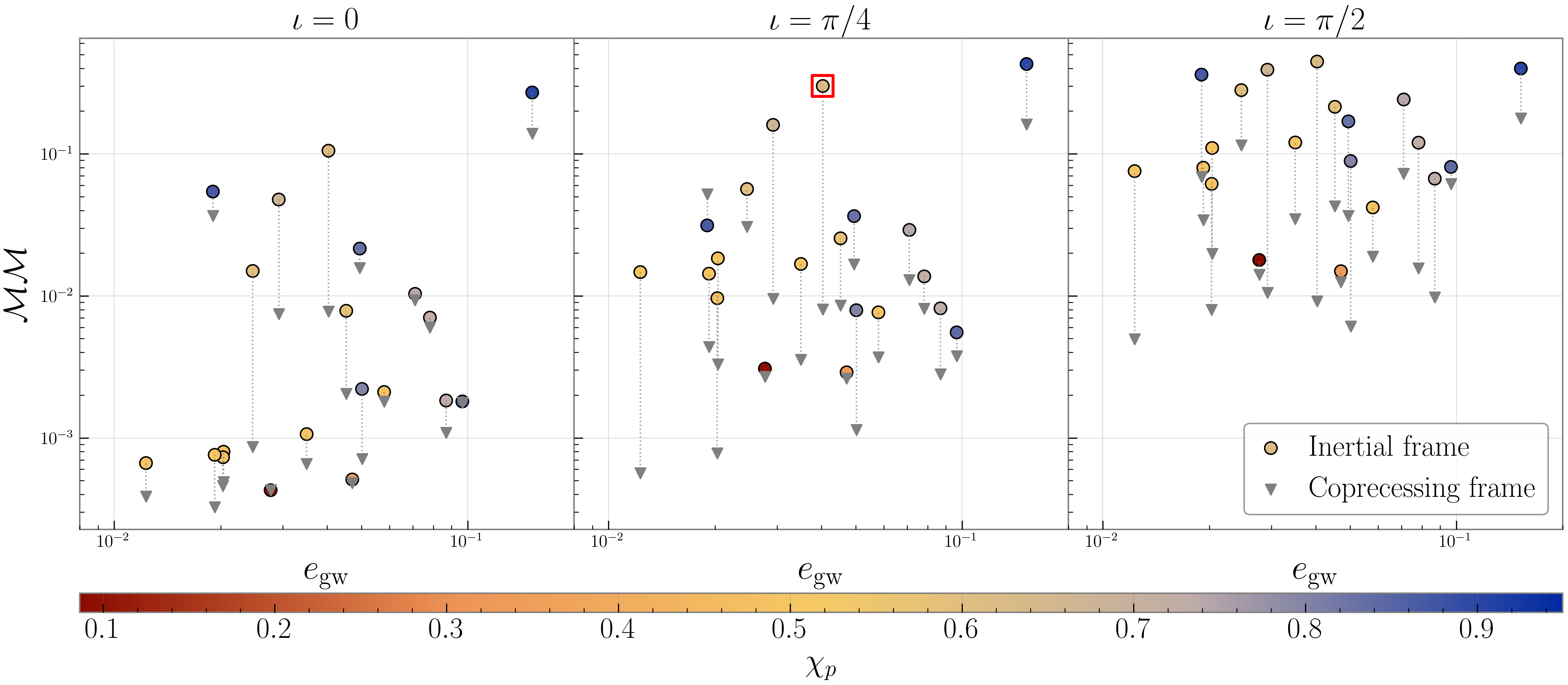}
    \caption{Mismatch $\mathcal{M}\mathcal{M}$ between the NR simulations from Tab.~\ref{tab:NRSimsParams} and the best-fitting \SEOB\,waveform, for $\iota=0$ (left), $\iota=\pi/4$ (middle), and $\iota=\pi/2$ (right). 
    Results are plotted as a function of the NR eccentricity $e_{\text{gw}}$ and coloured by the value of the precessing spin $\chi_p$ at the simulation reference time.
    Mismatches between \SEOB\, and NR in the inertial frame are shown by circles, while the equivalent mismatches for the same NR simulations transformed into the coprecessing frame are shown with grey triangles; mismatches in the two frames for the same simulation are joined by a grey dotted line. 
    The red box marks simulation \texttt{SXS:BBH:3714}, examined further in Fig.~\ref{fig:3714strain}.
    The eccentricity $e_{\text{gw}}$ is computed from the NR waveforms, as detailed in Sec.~\ref{subsec:NRSims}, and it is the same in the two frames.}
    \label{fig:strainmismatches}
\end{figure*}

In this section, we explore the degree to which an eccentric and precessing waveform, when transformed into the coprecessing frame, resembles the waveform produced by an eccentric but spin-aligned system. 
This resemblance forms the basis of all precessing waveforms of the EOB~\cite{Pan:2013rra, Ramos-Buades:2023ehm,Gamba:2021ydi} and Phenom~\cite{Schmidt:2012rh, Hamilton:2025xru} families. 
This is in contrast to precessing surrogate models \cite{Blackman:2017pcm}, which also use the coprecessing frame but do not impose any assumption that the coprecessing waveform resembles an approximately non-precessing one, only that it is smoothly-varying (see Sec.~\ref{sec:BuildingReducedBases}).
Modern Phenom and EOB models do not assume that the coprecessing waveform is \emph{exactly} a physical spin-aligned waveform \cite{Hamilton:2025xru,Ramos-Buades:2023ehm}.
Rather, they incorporate physical precession effects in the coprecessing frame such as asymmetries between positive and negative $m$-modes \cite{Thompson:2023ase, Estelles:2025zah} and changes to the ringdown frequencies and final spin \cite{Hamilton:2023znn} to increase waveform faithfulness.
Here, we explore how the addition of eccentricity affects this resemblance.

Following Sec.~\ref{subsec:MismatchProcedure} we compute mismatches and best-fitting \SEOB\, waveforms against the NR simulations of Table~\ref{tab:NRSimsParams}.
We do so both before and after transforming the NR waveforms into the coprecessing frame. 
We compare the full waveform strains rather than the individual modes, as a primary effect of precession is mode-mixing between modes with the same $m$-value, which would not be captured by only considering a single mode at a time. 
Computing strains requires a choice for the inclination $\iota$ which controls the relative strength of different modes.
We present results for $\iota\in\lbrace0, \pi/4, \pi/2\rbrace$ in Fig.~\ref{fig:strainmismatches}.

For virtually all\footnote{We find a single case across all simulations and inclinations for which the mismatch depreciates with the introduction of the coprecessing frame, simulation \texttt{SXS:BBH:3715} at inclination of $\iota=\pi/4$. Upon further investigation, this is due to \SEOB~ mismodeling certain persistent amplitude modulations close to merger.
Upon removing the final $200\,M$ prior to merger (around $3.5$ cycles), the mismatches improve by an order of magnitude, $1.7\times10^{-2}$ in the inertial frame and dropping to $1.5\times10^{-2}$ in the coprecessing frame.} inclinations and simulations, the mismatch between NR and \SEOB\, decreases when the NR simulation is transformed into the coprecessing frame, thus quantifying the conclusions of Fig.~\ref{fig:coprec_frame}. 
This confirms that the transformation does indeed suppress the dominant effects of precession, and we are left with a waveform which more closely resembles an eccentric aligned-spin one. 
As expected, inertial-frame mismatches increase with inclination as the relative importance of the higher-order modes, and hence the amount of mode-mixing, increases. 
Coprecessing-frame mismatches are improved for all inclinations, however the improvement is modest when compared with the requirements of waveform modeling~\cite{Chatziioannou:2017tdw,Purrer:2019jcp,Toubiana:2024car,Thompson:2025hhc}.
For $\iota=0$, the lowest-spin simulations have either low mismatches in the inertial frame, or benefit from order-of-magnitude improvements in the coprecessing frame. 
However, simulations with larger spin, $\chip\sim 0.7$, have mismatches $\sim 0.1-0.01$ in either frame.
The situation is worse for $\iota=\pi/2$ where coprecessing-frame mismatches do not drop appreciably below $0.01$ for any simulation.

\begin{figure*}[]
    \centering
    \includegraphics[width=\textwidth]{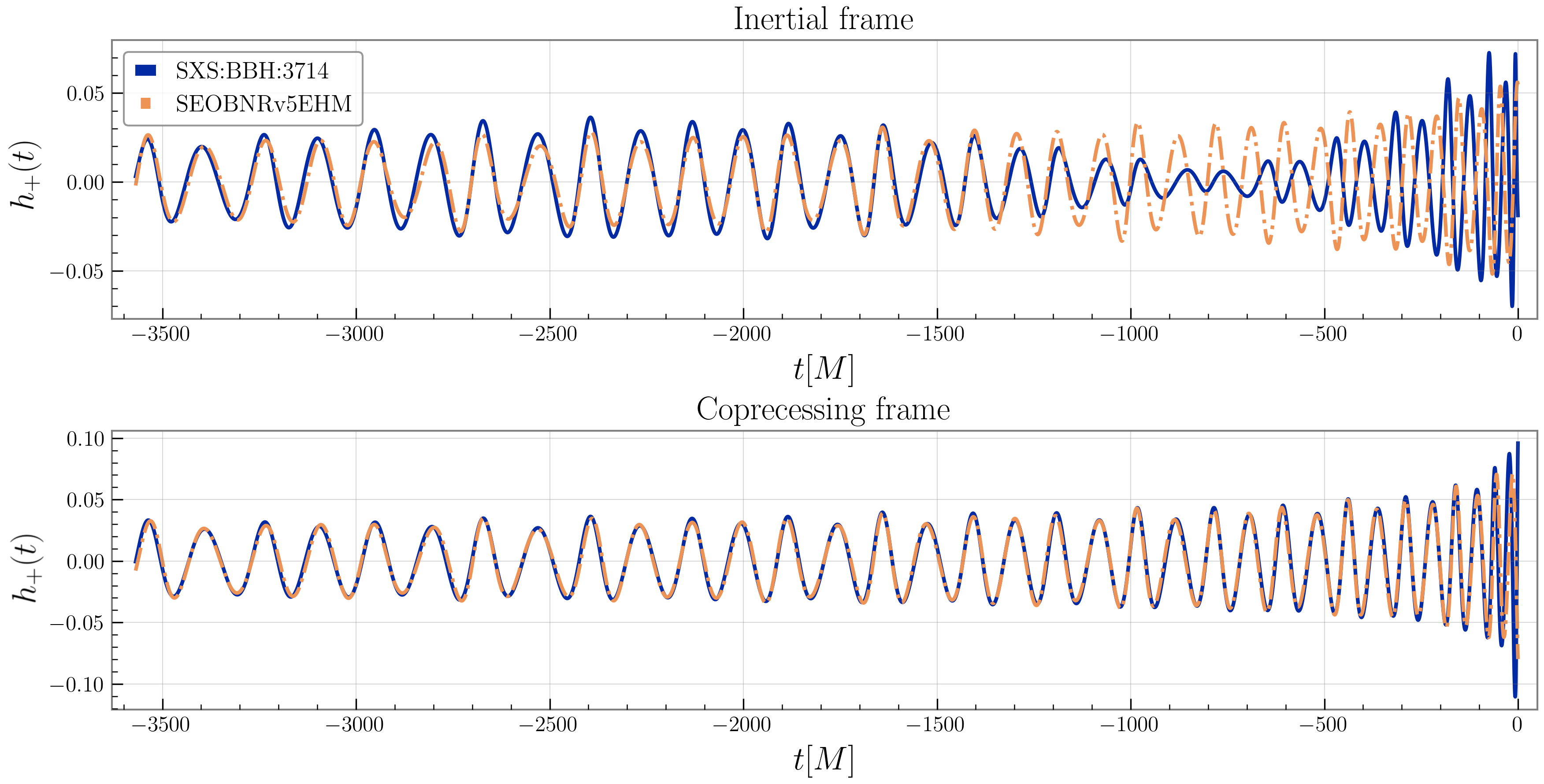}
    \caption{NR and best-fit \SEOB\, waveforms (plus polarization) as a function of time in the inertial (top panel) and coprecessing (bottom panel) frames for simulation \texttt{SXS:BBH:3714} and an inclination of $\iota=\pi/4$. This case is marked with a red box in Fig.~\ref{fig:strainmismatches} and demonstrates one of the largest mismatch reductions among all cases considered.}
    \label{fig:3714strain}
\end{figure*}
To gain more intuition for the mismatch improvement, in Fig.~\ref{fig:3714strain} we plot the NR and best-fit \SEOB\, waveforms in both frames for the simulation \texttt{SXS:BBH:3714} and $\iota=\pi/4$.
This represents one of the largest mismatch decreases due to coprecessing transformation from $\sim 0.300$ to $\sim 0.008$; the case is also marked by a red box in Fig.~\ref{fig:strainmismatches}. 
The large precession modulation after $t=-1500\,M$ drives the disagreement between the waveforms' amplitude and phase in the inertial frame.
This morphology disappears in the coprecessing frame and the waveforms agree in both amplitude and phase (including, crucially, an amplitude modulation due to eccentricity).

\begin{figure}[]
    \centering
    \includegraphics[width=\columnwidth]{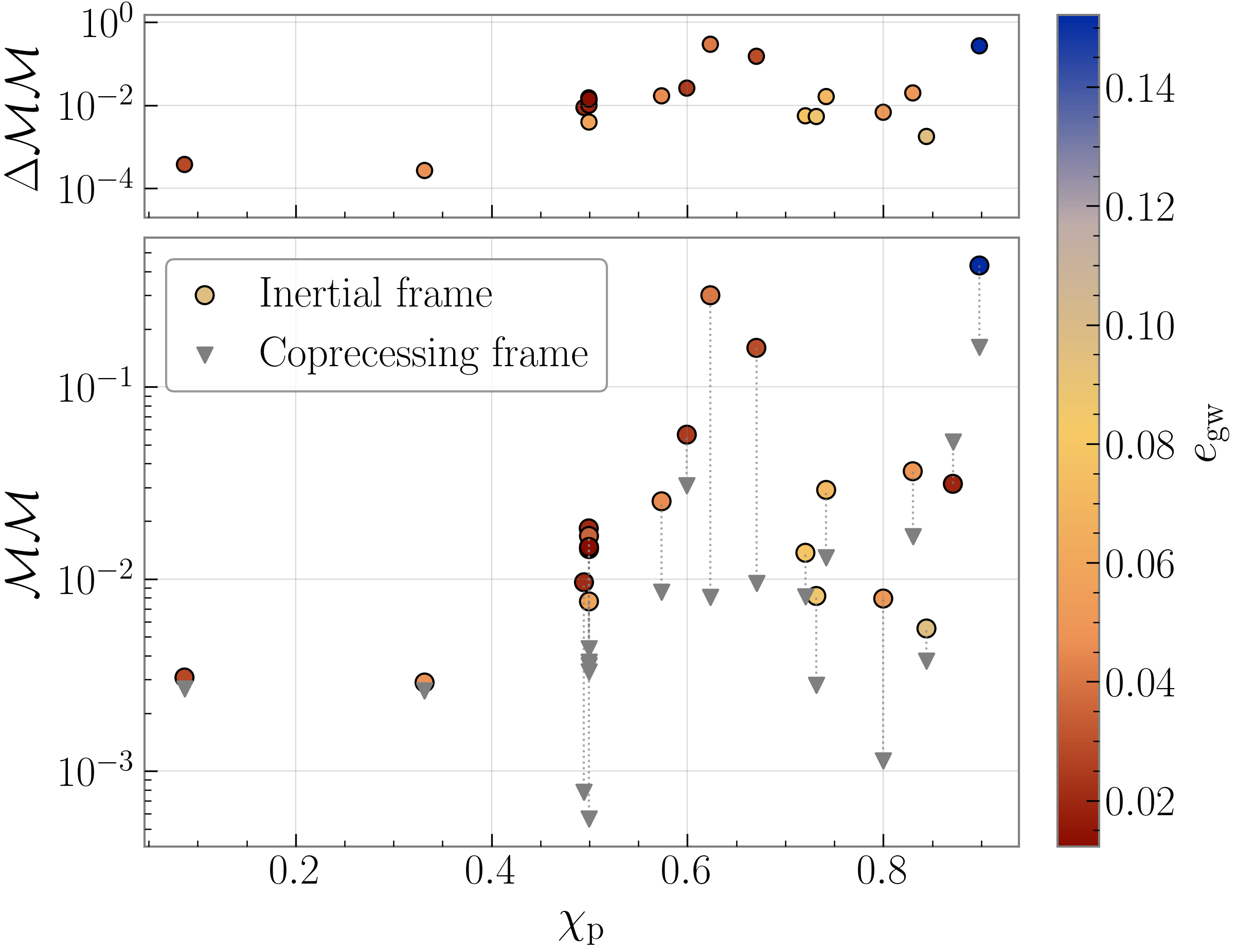}
    \caption{Mismatches $\mathcal{M}\mathcal{M}$ in both frames (main panel) and mismatch difference between the inertial and coprecessing frame (top panel) as a function of $\chi_{\text{p}}$ (at the simulation reference time) and for $\iota=\pi/4$. These are the same data as the middle panel of Fig.~\ref{fig:strainmismatches}, now plotted as a function of the precessing spin.}
    \label{fig:strainmismatches_chip}
\end{figure}

We study the impact of the amount of precession present in Fig.~\ref{fig:strainmismatches_chip}, where we plot mismatches in both frames as well as their difference as a function of the effective precession spin $\chi_{\text{p}}$ for $\iota=\pi/4$. 
In general, mismatch values in both frames tend to increase with larger $\chi_{\text{p}}$ values,
indicating that precession effects are a key driver for the differences between the waveforms.
This trend is also true in the coprecessing frame, which indicates that the transformation does not completely remove the effects of precession, as expected from, e.g., mode asymmetry.
Also correlated with $\chip$ is the improvement in mismatch when transforming to the coprecessing frame, suggesting that the transformation is still capturing most of precession and can aid in waveform modeling.
We obtain qualitatively similar conclusions for the other inclination values. 

\begin{figure}[]
    \centering
        \includegraphics[width=\columnwidth]{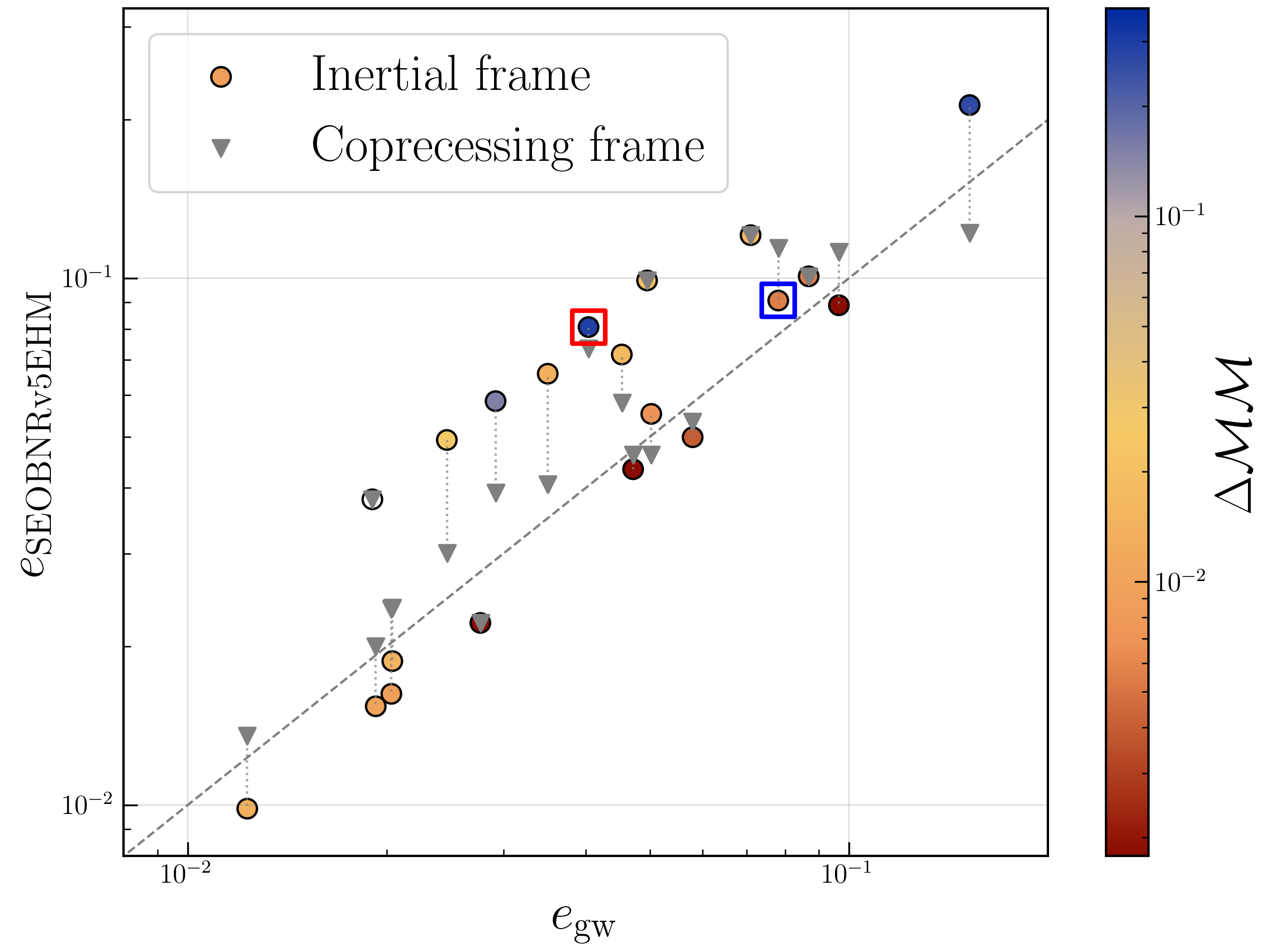}
        \caption{Eccentricities of the NR simulation ($e_{\text{NR}}$) and the best-fitting \SEOB\, waveform ($e_{\text{SEOBNRv5EHM}}$), in both the inertial (coloured circles) and coprecessing (grey triangles) frames for $\iota=\pi/4$. The circles are coloured by the value of the mismatch difference between the inertial and coprecessing frames. The red box marks simulation \texttt{SXS:BBH:3714}, examined further in Fig.~\ref{fig:3714strain}, where the eccentricity remains similar despite a large mismatch improvement. The blue box marks a case (\texttt{SXS:BBH:3711}), where the eccentricity becomes more different than the NR one after the coprecessing transformation.}
        \label{fig:ecc_comparison}
\end{figure}


The mismatch calculation involves maximization over several parameters, see Sec.~\ref{subsec:MismatchProcedure}.
Traditionally, this has led to the distinction between the \emph{mismatch} (maximize over phase and time) and the \emph{fitting factor} (maximize over all parameters)~\cite{Buonanno:2009zt}.
The former quantifies the accuracy of a waveform for parameter estimation, namely its ability to model some data with the correct parameters.
The latter is relevant to detection, as it allows a model to ``distort'' its parameters in order to model the data.
The difference between the true and the best-fit physical parameters of the fitting factor are then related to the amount of systematic bias induced by waveform mismodeling.
A bigger improvement between the mismatch and the fitting factor signals larger systematics.

Systematics are subtly different for eccentricity.
Even though models have some input eccentricity parameter, the final quoted eccentricity value is extracted from the waveform itself, see Sec.~\ref{subsec:NRSims}.
Therefore in principle the more similar two waveforms are, i.e., a higher fitting factor in the language of the previous paragraph, the more similar their eccentricities should be.
In practice, however, the quoted eccentricity is extracted from only two binary orbits and further varies with time, so a reference time is chosen.
So the degree to which eccentricity is systematically biased will depend on whether the waveforms agree during the critical cycles where it is defined.
This simplifies the issue of systematics in eccentricity for observed data as long as the eccentricity is defined from the observed signal, and not an earlier time or frequency.
This is because models result in similar raw waveforms, albeit with potentially different input parameters, c.f., Fig.~1 of Ref.~\cite{Bini:2026kwz} for GW231123~\cite{LIGOScientific:2025rsn} which has large parameter systematics but the reconstructed waveforms are still similar.

We explore the eccentricities of the different waveforms in Fig.~\ref{fig:ecc_comparison}, where both NR and \SEOB\, eccentricities are defined from the waveform per Sec.~\ref{subsec:NRSims}. 
Since we minimize the mismatch between NR and \SEOB\, separately in the coprecessing and inertial frames, we have two corresponding best-fit eccentricities. 
We show results for $\iota=\pi/4$, but we obtain qualitatively similar results for other inclination values. 
The NR and \SEOB\, eccentricities show good agreement, which further improves for most cases in the coprecessing frame.
This is broadly consistent with Fig.~\ref{fig:strainmismatches} and suggests that the waveform agreement is improved in the first few cycles where eccentricity is defined: mismatches are lower in the coprecessing frame, therefore the waveforms are more similar and their eccentricities are correspondingly more similar.
There are a handful of exceptions, however, where the eccentricity agreement is worse, e.g., \texttt{SXS:BBH:3711} marked with a blue box, even though the mismatch improves.
In these cases, the mismatch improves due to better waveform agreement in the later inspiral stages, while the agreement slightly deteriorates in the first few cycles and the extracted eccentricity is correspondingly affected.
An interesting case is again \texttt{SXS:BBH:3714} (red box) that has a large mismatch improvement but minimal eccentricity change between the two frames. 
This can be understood by inspection of Fig.~\ref{fig:3714strain}: the coprecessing transformation massively improves waveform agreement during the late inspiral but leaves the first 2 orbits largely unaffected.



\section{Smoothness of Coprecessing Frame Waveform for Surrogate modeling}
\label{sec:BuildingReducedBases}

So far, we have shown that the coprecessing frame transformation effectively reduces the waveform's complexity even in the presence of eccentricity. 
Analytic models utilize this transformation to model the coprecessing waveform as that of an equivalent spin-aligned system (plus corrections~\cite{Thompson:2023ase, Estelles:2025zah, Hamilton:2023znn, Hamilton:2025xru}).
A complementary approach is that of full surrogate modeling that makes no assumptions about the physical interpretation of the coprecessing frame waveform.
Rather, surrogates to NR simulations treat the coprecessing frame as a \emph{convenient} transformation that simplifies the waveform and aids interpolation.
In this section, we explore the coprecessing frame in the context of these models, following Refs.~\cite{Blackman:2017pcm,Varma:2019csw}.

Surrogate modeling starts with a preexisting set of waveforms (for example an NR catalog~\cite{Scheel:2025jct}) and creates an interpolant that can be evaluated at any point in the parameter space.
Waveforms are broken down into components $X(t;\vec{\Theta})$ such as the waveform amplitude and phase, where $\vec{\Theta}$ again represents the intrinsic parameters.  
Across the parameter space of intrinsic parameters that we wish to model $\mathcal{T}$, $X(t;\vec{\Theta})$ is represented with some orthonormal linear basis $B_{n}=\lbrace e^{i}(t)\rbrace^{n}_{i=1}$ as
\begin{equation}
    X(t;\vec{\Theta}) \approx \sum_{i=1}^n c_{i}(\vec{\Theta})e^{i}(t)\,,
    \label{eq:basis}
\end{equation}
for $\vec{\Theta} \in \mathcal{T}$, where $c_{i}(\vec{\Theta})$ are the basis coefficients that encode the dependence on the binary parameters. 
The orthonormal basis elements are functions of time $e^{i}(t)$ that can be found using a greedy algorithm, in which basis elements are added successively from the preexisting waveforms until the basis reaches a predetermined accuracy. 
The accuracy is defined using a mean squared projection error among the preexisting waveforms,
\begin{equation}
E_{n}(\vec{\Theta}) = \sum_{t} \left(X(t;\vec{\Theta}) - \sum_{i=1}^{n} c_{i}(\vec{\Theta})e^{i}(t)\right)^2\,.
\label{eq:basisprojection}
\end{equation}
With each successive iteration of the greedy algorithm, the waveform with the maximum projection error is orthonormalized and added to the basis. 
The algorithm stops when either of the following two conditions is satisfied: the maximum projection error is below the user-defined tolerance; or the waveform with the maximum projection error is already inside the basis.
With the basis in hand, the next step is to construct an empirical interpolant to compress the number of time nodes (or alternatively parameterized by anomaly for eccentric systems~\cite{Nee:2025nmh,Maurya:2025shc}) and fit the reduced basis coefficients at the empirical nodes across the parameter space.

\begin{figure}[]
    \centering
    \includegraphics[width=\columnwidth]{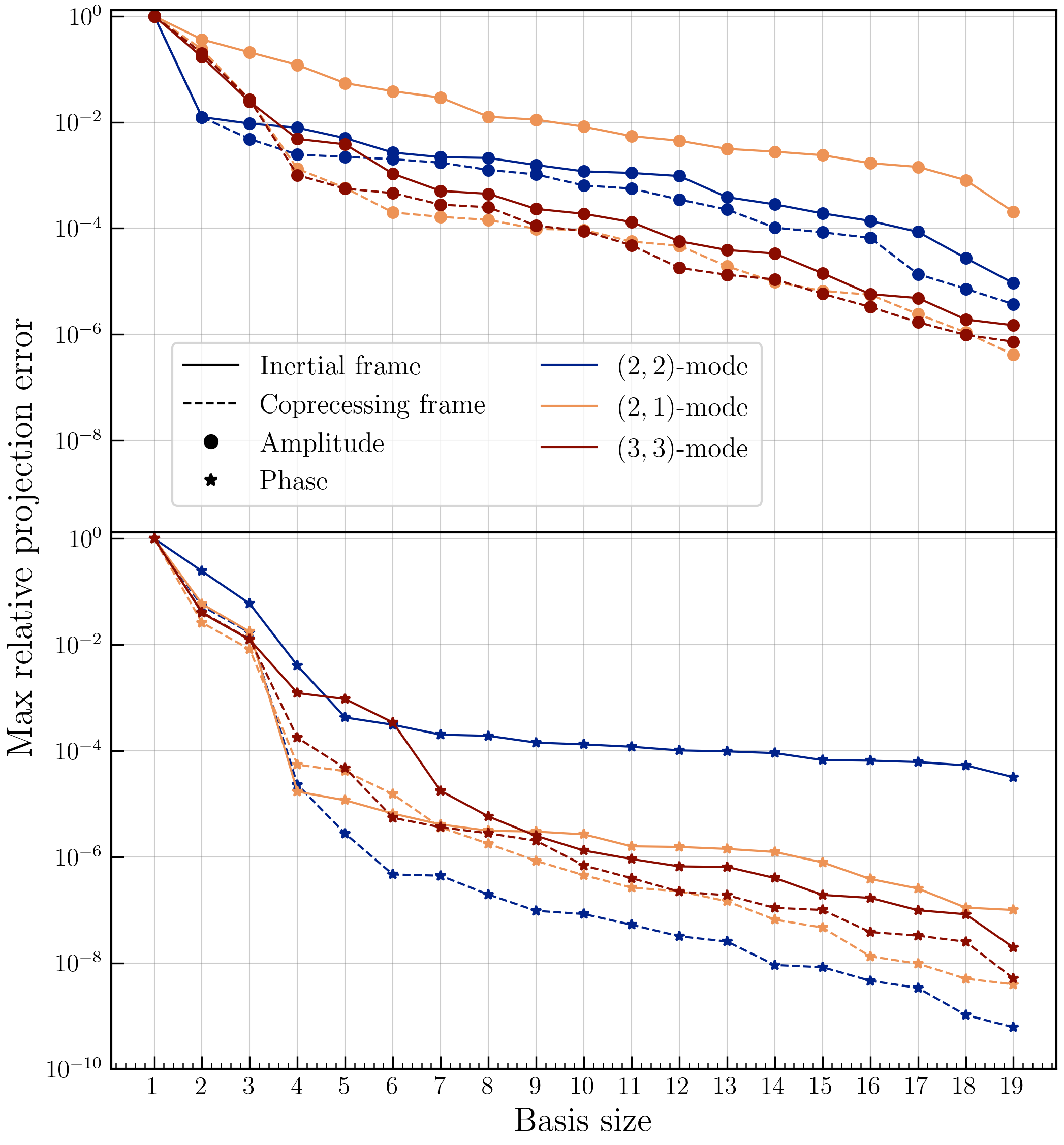}
    \caption{Reduced basis error as a function of basis size, constructed in both the inertial (solid lines) and coprecessing (dashed lines) frames, for the amplitude (top plot, circle markers) and phase (bottom plot, star markers) of the $(2,2)$ (blue) $(2,1)$ (orange) and $(3,3)$ (red) modes.}
    \label{fig:basiserrors}
\end{figure}

Here, we compute the reduced basis but neglect the coefficients due to the very low number of waveforms. 
We also choose a parameterization in time rather than mean anomaly due to the short length of several NR waveforms.
The basis construction proceeds as follows.  
First, we shift the waveforms such that the peak is aligned as in Eq.~\eqref{eq:hamp}. 
Then, we remove the postmerger part of each waveform, after the peak amplitude. 
As the length of the components covered by the basis are limited by the shortest waveform, we also remove all waveforms with a length $<2000M$, which leaves us with 19 waveforms. 
Finally, we make bases for the amplitude and phase of the modes $h_{\ell,m}$ where $(\ell,m) \in \lbrace(2,2), (2,1), (3,3)\rbrace$. 

We construct bases in both the inertial and coprecessing frames and plot the evolution of the reduced basis error from Eq.~\eqref{eq:basisprojection} as a function of the basis size in Fig.~\ref{fig:basiserrors}.
For all surrogate components and waveform modes, the error on average reduces faster in the coprecessing frame than the inertial one as basis elements are added (compare solid to dashed lines of same color).
Though the errors oscillate especially for the first few basis sizes, the average trend of decreasing errors with more bases and lower errors in the coprecessing frame is consistent. 
This again confirms that the waveforms as represented in the coprecessing frame vary more smoothly across the parameter space, so the waveform set can be represented more accurately by a smaller number of waveforms.

The largest improvement between the inertial and coprecessing frame occurs for the amplitude of the $(2,1)$-mode (top plot, orange solid and dashed lines), and the phase of the $(2,2)$-mode (bottom plot, blue solid and dashed lines). 
Regarding the $(2,1)$-mode amplitude, this matches the expectation that the imprint of precession is more prominent in this mode due to mode-mixing, for example as shown by the second row of Fig.~\ref{fig:coprec_frame}. 
Therefore removing the precession modulation simplifies the $(2,1)$ amplitude, meaning it can be better modeled with a given number of basis elements.
Regarding the $(2,2)$-phase, its accuracy is the dominant contribution to modeling, since the $(2,2)$-mode dominates most BBH signals, as it is the waveform phase, not amplitude, which primarily affects the mismatch in Eq.~\eqref{eq:mismatch}. 
Therefore, the improved accuracy shown especially for this component in Fig.~\ref{fig:basiserrors} implies that when building surrogate models of eccentric and precessing systems, the coprecessing frame representation is key for optimal accuracy.

\section{Conclusions}
\label{sec:conclusions}

We have investigated the impact of orbital eccentricity on the coprecessing frame, a time-dependent transformation that forms the basis of all GW models for precessing binaries.
Using 20 eccentric and precessing NR simulations from the latest SXS catalog, we have confirmed that the coprecessing frame simplifies the waveform morphology by removing amplitude modulations and mode-mixing.
We have confirmed that eccentric, non-precessing waveforms are left unaltered by the coprecessing frame transformation.

We have further quantitatively assessed the extent to which the coprecessing frame facilitates waveform modeling.
Regarding analytical models, we find that mismatches between \SEOB\, (a waveform that includes eccentricity but no precession) and the NR simulations universally improve when the latter are transformed into the coprecessing frame.
The improvement further correlates with the amount of precessing spin $\chip$.
However, mismatches remain large compared to the needs of unbiased GW modeling, $\gtrsim 0.01$ for inclination $\pi/2$.
Even though the transformation suppresses the dominant effects of precession in the presence of eccentricity, corrections to the coprecessing frame waveform remain essential.

Regarding direct surrogates to NR, the coprecessing frame aids modeling by reducing the complexity of the parameter space and waveform morphology.
Reduced bases constructed in the coprecessing frame reach a given accuracy with fewer basis elements than the inertial frame one.
The largest improvements are seen in the $(2,1)$ mode amplitude due to removal of mode-mixing and the $(2,2)$ mode phase which is particularly important for modeling.
Though our study is based on only $20$ publicly available simulations, our analysis suggests that the coprecessing frame will remain a cornerstone of precessing BBH modeling in the presence of eccentricity.

\acknowledgments
The authors thank Md Arif Shaikh for helpful discussions, and Geraint Pratten for useful comments on the manuscript.
LMT is supported by NSF MPS-Gravity Awards 2207758 and
2513294, and by NSF Grant 2309200.
KC and TK were supported by NSF Grant PHY-2409001 and the Sherman Fairchild Foundation at Caltech.
SJ was supported by the LIGO Laboratory Summer Undergraduate Research Fellowship program (LIGO SURF), and the California Institute of Technology Student Faculty Programs.
This material is based upon work supported by the National Science Foundation under Grants No. PHY-2407742, No. PHY-2207342, and No. OAC-2513338 at Cornell. Any opinions, findings, and conclusions or recommendations expressed in this material are those of the author(s) and do not necessarily reflect the views of the National Science Foundation. This work was supported by the Sherman Fairchild Foundation at Cornell.
The authors are grateful for computational resources provided by the LIGO Laboratory and supported by National Science Foundation Grants PHY-0757058 and PHY-0823459.
This material is based upon work supported by NSF's LIGO Laboratory which is a major facility fully funded by the National Science Foundation.
\\
\subsection*{Software Used}
\texttt{sxs} \cite{Boylesxs}, \texttt{scri} \cite{Boylescri}, \texttt{gw\_eccentricity} \cite{Shaikh:2023ypz,Shaikh:2025tae}, \texttt{PySEOBNR} \cite{Mihaylov:2023bkc}, \texttt{LALSimulation} \cite{lalsuite,swiglal}, \texttt{SciPy} \cite{2020SciPy-NMeth}, \texttt{HTCondor} \cite{Thain2005}

\bibliography{main}
\end{document}